\documentclass[journal=jacsat,manuscript=article]{achemso}

\usepackage[version=3]{mhchem} 

\author{Mikhail Fonin}
\affiliation[Universit\"at Konstanz]{Fachbereich Physik, Univiersit\"at Konstanz, 78457 Konstanz, Germany}
\email{mikhail.fonin@uni-konstanz.de}

\author{Muriel Sicot}
\affiliation[Universit\"at Konstanz]{Fachbereich Physik, Univiersit\"at Konstanz, 78457 Konstanz, Germany}

\author{Ole Zander}
\affiliation[Universit\"at Konstanz]{Fachbereich Physik, Univiersit\"at Konstanz, 78457 Konstanz, Germany}

\author{Samuel Bouvron}
\affiliation[Universit\"at Konstanz]{Fachbereich Physik, Univiersit\"at Konstanz, 78457 Konstanz, Germany}

\author{Philipp Leicht}
\affiliation[Universit\"at Konstanz]{Fachbereich Physik, Univiersit\"at Konstanz, 78457 Konstanz, Germany}

\author{Ulrich R\"udiger}
\affiliation[Universit\"at Konstanz]{Fachbereich Physik, Univiersit\"at Konstanz, 78457 Konstanz, Germany}

\author{Martin Weser}
\affiliation[Fritz-Haber Institut]{Fritz-Haber Institut der Max-Planck Gesellschaft, 14195 Berlin, Germany}

\author{Yuriy S. Dedkov}
\affiliation[Fritz-Haber Institut]{Fritz-Haber Institut der Max-Planck Gesellschaft, 14195 Berlin, Germany}

\author{Karsten Horn}
\affiliation[Fritz-Haber Institut]{Fritz-Haber Institut der Max-Planck Gesellschaft, 14195 Berlin, Germany}

\title{Spatial corrugation and bonding of single layer graphene on Rh(111)}

\begin{document}
\begin{abstract}
Topographic scanning tunneling microscopy (STM) images of epitaxial single layer graphene on the Rh(111) surface reveal that extended single crystalline graphene domains are produced without any defects on a large scale. High resolution imaging shows that the moir\'e structure resulting from the lattice mismatch between the Rh(111) substrate and graphene is highly corrugated, containing regions of an additional spatial modulation in the moir\'e supercell compared with those previously reported for graphene on Ir(111) or graphene on Ru(0001). These areas, which correspond to the "bridge" regions of the moir\'e structure appear as depressions in STM images indicating a strong orbital hybridization between the graphene layer and the metallic substrate. Valence-band photoemission confirms the strong hybridization between graphene and Rh(111) which leads to the pronounced corrugation of the graphene layer. Our findings underline the importance of considering substrate effects in epitaxially grown graphene layers for the design of graphene-based nanoscale systems.
\end{abstract}


The exceptional transport properties of graphene, a two-dimensional honeycomb lattice of $sp^2$ bonded carbon atoms~\cite{Geim:2007}, make it a promising material for applications in microelectronics~\cite{Novoselov:2005,Morozov:2008} and sensing~\cite{Schedin:2007}. This has recently led to a revival of interest in graphene growth on transition metal surfaces which might be an alternative to micromechanical cleavage for producing macroscopic graphene films~\cite{Kim:2009}. The most studied substrates include Ni(111)~\cite{Gamo:1997,Dedkov:2008a,Dedkov:2008b}, Ru(0001)~\cite{Himpsel:1982,Marchini:2007,Martoccia:2008,Sutter:2008,Praga:2008,Wang:2008}, Ir(111)~\cite{Gall:2000,Gall:2004,NDiaye:2006,Coraux:2008,Coraux:2009}, and Pt(111)~\cite{Lyon:1967,Hu:1987,Land:1992,Enachescu:1999,Sasaki:2000}. The latter three systems show a considerable mismatch between the metal lattice and the graphene lattice, which together with strong C-C $\sigma$-bonding, causes the formation of moir\'e superstructures accompanied by a periodic corrugation of the graphene layer~\cite{Wintterlin:2009}. Atomically resolved scanning tunneling microscopy (STM) investigations of graphene/Ru(0001)~\cite{Marchini:2007,Wintterlin:2009} revealed a strong spatial variation of the STM contrast, which was associated with the stronger chemical bonding compared with other mismatched graphene/transition-metal interfaces. As for the graphene/Rh(111) interface, which is believed to represent a transition case from strongly bonded and highly corrugated graphene/Ru(0001) to the weakly bonded graphene/Ir(111) system~\cite{Preobrajenski:2008}, little information is available about the exact atomic arrangement of the graphene layer with respect to the substrate. The investigation of the atomic structure of graphene/Rh(111) may provide an important insight into the interaction of graphene with underlying transition metal substrates as well as deliver valuable information in potential applications, such as a possible template for the growth of metallic nanocluster arrays~\cite{Sicot:2010}. To this end, we address the structure and electronic properties by STM and photoemission spectroscopy (PES) of epitaxial single layer graphene on Rh(111). High-resolution imaging of the graphene surface shows that the moir\'e structure has a distinctly different atomic configuration compared with the previously reported structures for graphene on Ir(111)~\cite{NDiaye:2006} or graphene on Ru(0001)~\cite{Marchini:2007,Martoccia:2008,Wang:2008} as well as for the \textit{h}-BN "nanomesh" on Rh(111)~\cite{Corso:2004}. Surface areas showing the strongest interaction between the graphene layer and the metallic substrate are confined to the "bridge" regions of the moir\'e superstructure. Valence-band photoemission confirms the strong hybridization between graphene and Rh(111) which leads to the pronounced corrugation of the graphene layer.

All STM experiments were carried out in an ultra-high vacuum (UHV) system (base pressure 1$\times$10$^{-10}$\,mbar) equipped with an Omicron variable temperature scanning tunneling microscope. All STM measurements were performed in the constant-current-mode at room temperature using electrochemically etched polycrystalline tungsten tips cleaned in UHV by flash-annealing. The sign of the bias voltage corresponds to the  voltage applied to the sample. The Rh(111) single crystal (MaTeck GmbH, purity 99.99\,\%) was cleaned by repeated cycles of Ar$^+$ sputtering at room temperature [$p$(Ar)\,=\,5$\times$10$^{-6}$\,mbar, 1.2\,keV], flash-annealing to about 1500\,K, annealing in an oxygen atmosphere [$p$(O$_2$)\,=\,1$\times$10$^{-7}$\,mbar] at about 1000\,K, and subsequent flash-annealing in UHV to about 1500\,K. The quality of the Rh(111) surface was subsequently checked by STM, LEED, and core-level as well as valence band PES. Graphene layers were prepared by thermal decomposition of propene gas [$p$(C$_3$H$_6$)\,=\,3$\times$10$^{-8}$\,mbar] at 900-1100\,K. Identical samples were grown for photoemission experiments. The presented valence band photoemission as well as core-level PES measurements were performed at the UE56/2-PGM-1 beamline at BESSY (Berlin). The PES spectra were collected with a PHOIBOS\,100 energy analyzer while the sample was placed on a 6-axis manipulator (3 translation and 3 rotation axis). The energy/angular resolution was set to 80\,meV/0.2$^\circ$.

\begin{figure*}[t]
  \vspace{5mm}
    \begin{center}
 \includegraphics[width=130mm]{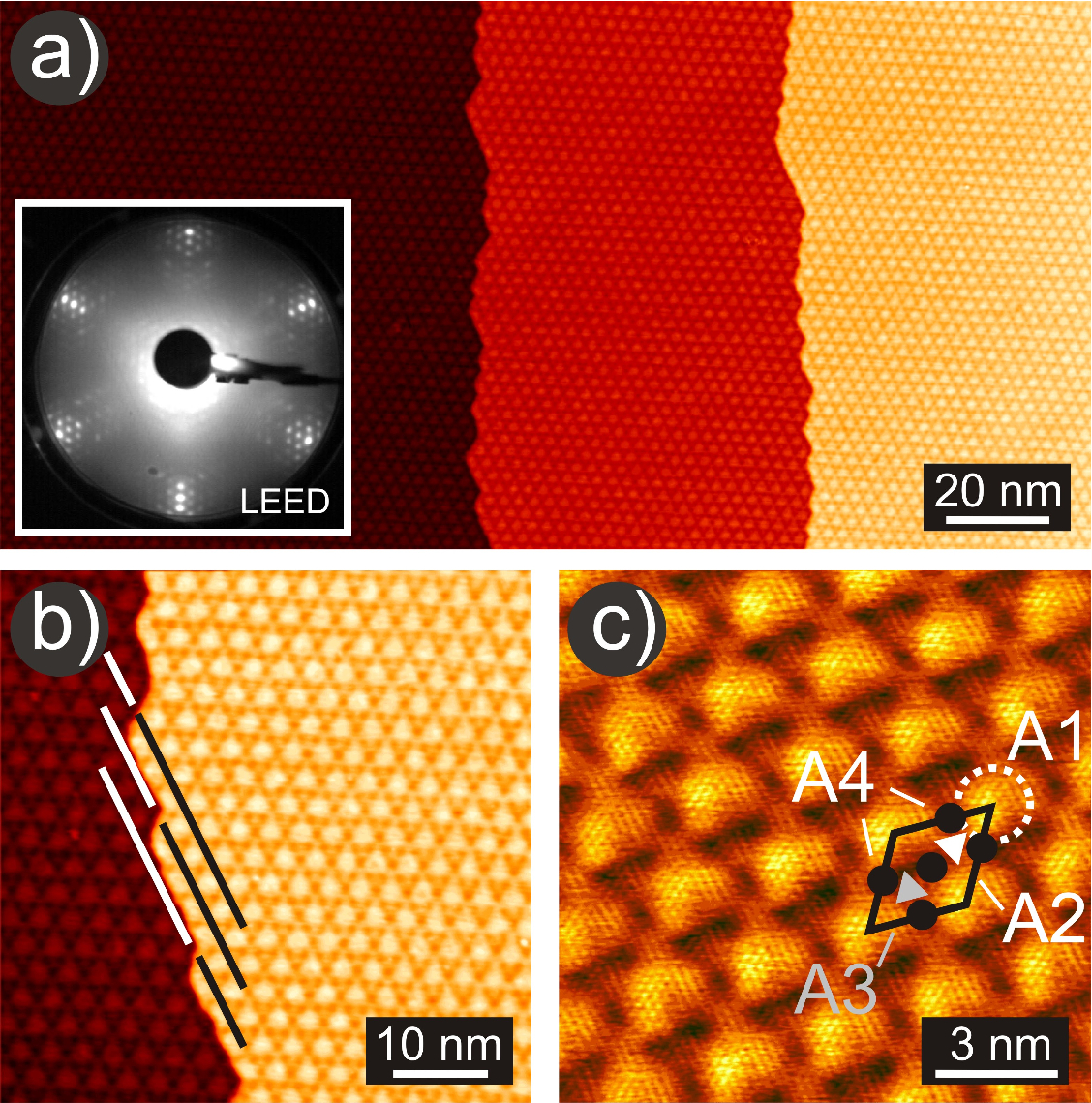}
    \end{center}
\caption{\label{figure01} (a) Large scale STM image of an epitaxial graphene layer on the Rh(111) surface. Tunneling parameters: U$_T$\,=\,1.1\,V; I$_T$\,=\,0.18\,nA. Inset: a LEED image of the graphene layer on Rh(111) taken at a primary electron energy of 58\,eV. (b) A higher magnification STM image of the graphene/Rh(111) moir\'e structure (U$_T$\,=\,1.1\,V; I$_T$\,=\,0.18\,nA). (c) Atomically-resolved image of the unit cell of the moir\'e structure (U$_T$\,=\,0.01\,V; I$_T$\,=\,2.5\,nA). The black rhombus outlines the supercell of the moir\'e structure. Main surface features are marked in the image: moir\'e maximum (A1) by a circle, the brighter half-cell (A2) by a white solid triangle, the darker half-cell (A3) by a grey solid triangle, and bridging black minima (A4) by black solid circles.}
\end{figure*}

In order to check the quality and continuity of the graphene layer as well as to provide structural details of single layer graphene at the atomic level, we performed \textit{in situ} STM measurements. An overview of a continuous graphene domain on the Rh(111) surface is shown in~\ref{figure01} (a). The graphene layer exhibits a highly ordered moir\'e structure without any visible defects even over large areas. \ref{figure01} (b) shows that the Rh steps are aligned along the main directions of the graphene layer which is attributed to the substantial reshaping of the Rh(111) surface during the graphene growth process in order to accommodate the periodicity and orientation of the graphene overlayer. This observation already indicates pronounced interactions between the graphene layer and the Rh surface. A typical low-energy electron diffraction (LEED) pattern of the graphene moir\'e on Rh(111) is presented in the inset of~\ref{figure01} (a). Qualitative analysis of STM and LEED images show that the close-packed directions of graphene and the unit cell vectors of the moir\'e are parallel to the close-packed $\langle1\bar{1}0\rangle$ directions of Rh(111). From the LEED images a periodicity of 2.90$\pm$0.1\,nm of the moir\'e superstructure on Rh(111) was calculated, which is in reasonable agreement with the average distance between the neighboring moir\'e features measured in STM images and corresponds roughly to 12 times the lattice constant of graphene and 11 times that of Rh(111). A higher magnification STM image of the graphene surface is shown in~\ref{figure01} (c) with the unit cell of the moir\'e superstructure marked by a rhombus. The apparent vertical corrugation of the graphene monolayer is measured to be in the range of 0.5\,-\,1.5\,\AA\ (peak-to-peak) depending on the tunneling conditions. Four distinctive regions corresponding to different apparent height levels can be distinguished within the moir\'e supercell: three maxima of different heights (A1, A2, A3) and black minima (A4). The most prominent maxima (A1) are surrounded by six black minima as well as by three less intense maxima A2 and A3. Each of the less intense maxima (A2, A3) is surrounded by three black minima (A4). The local atomic configuration of all observed features will be described below. 

\begin{figure*}[t]
  \vspace{5mm}
    \begin{center}
\includegraphics[width=120mm]{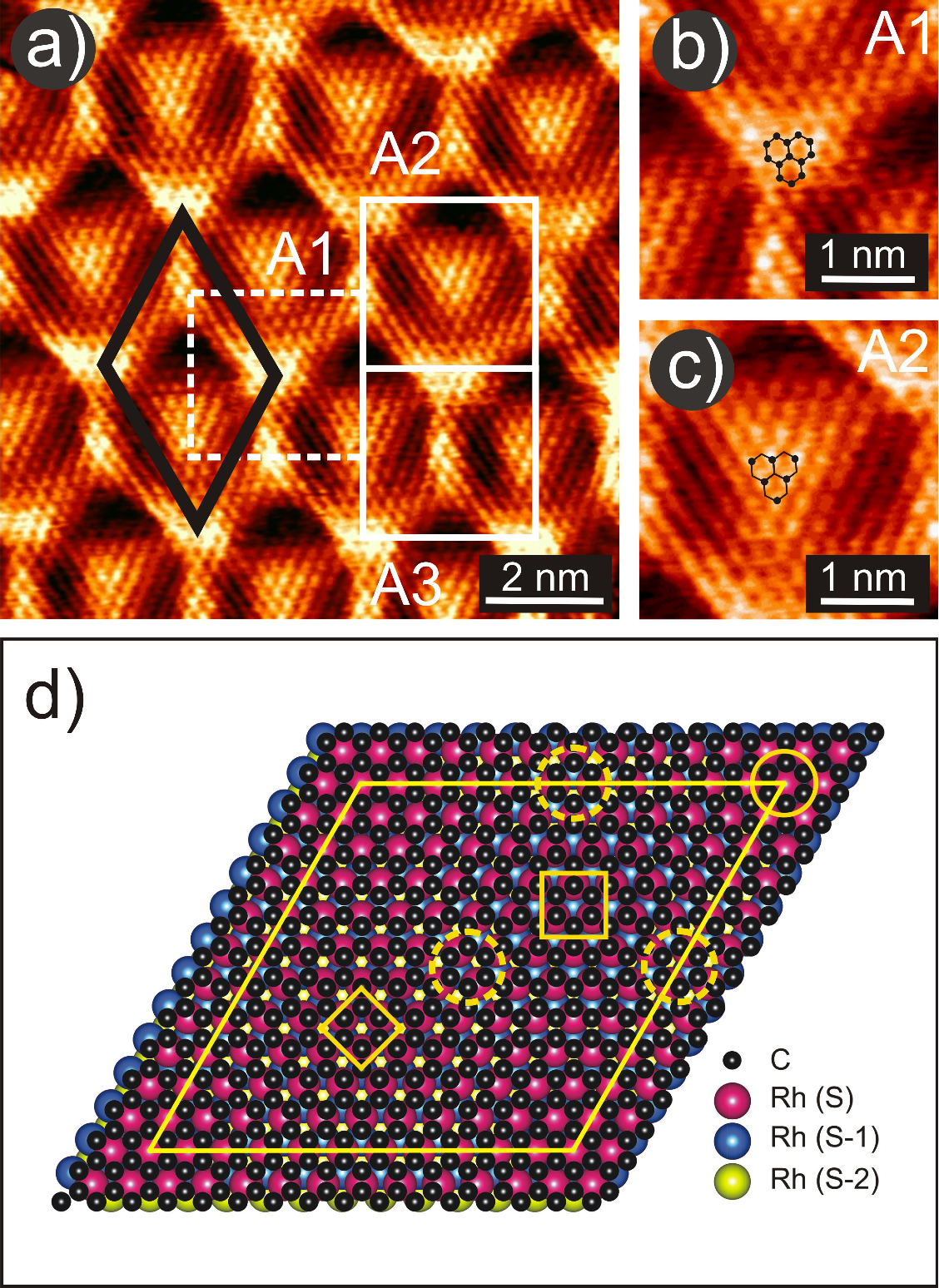}
    \end{center}
\caption{\label{figure02} (a) Atomically-resolved STM image of the moir\'e structure of the graphene layer on Rh(111). The rhombus outlines the supercell of the moir\'e superstructure with three different maxima marked by A1, A2 and A3. (b) Magnified view of the most prominent maximum (A1) which shows a honeycomb carbon lattice. (c) Magnified view of the A2 region of the moir\'e structure, showing only one carbon sublattice. Tunneling parameters for all images: U$_T$\,=\,0.02\,V; I$_T$\,=\,30\,nA. (d) Top-view of a simple ball model for the growth of graphene on Rh(111). Carbon atoms are black, first layer Rh atoms are red, second layer Rh atoms are blue, and third layer Rh atoms are yellow spheres. The yellow rhombus outlines the supercell of the moir\'e structure with four different positions: \textit{atop} (marked by a circle), \textit{top-fcc} (marked by a square), \textit{top-hcp} (marked by a rotated square), and "bridge-top" (marked by dashed circles).}
\end{figure*}

The high-resolution STM image in~\ref{figure02} (a) shows no atomic-scale defects, confirming a high crystalline quality of single layer graphene on the Rh(111) surface. The topographic features observed in~\ref{figure01} (c) can be clearly identified in this image. First, the most prominent maxima which are surrounded by six black minima as well as by three less prominent maxima correspond to the A1 maxima [see~\ref{figure02} (a)]. The difference between the two less intense maxima is now more prominent with one feature (A2) occupying an apparently larger surface area than the other one (A3). Each of the features is surrounded by three black minima A4. In most regions of the moir\'e supercell the graphene sublattice symmetry is broken and only one of the two carbon sublattices is imaged. We attribute this effect to the strong, covalent interaction between the graphene layer and the Rh(111) substrate. As an example of the symmetry breaking, magnified STM images of A1 and A2 features are presented in~\ref{figure02} (b) and (c), respectively. 

A more detailed description of the atomic configurations can be carried out on the basis of a simple ball model presented in~\ref{figure02} (d). Three highly-symmetric positions of carbon atoms with respect to the underlying Rh lattice can be assigned: \textit{atop} sites at the corners of the moir\'e unit cell, \textit{top-hcp} and \textit{top-fcc} sites in the center of the left and right half-cell of the moir\'e unit cell, respectively. In the STM image [~\ref{figure02} (a)] \textit{atop} regions appear as the bright maxima of the moir\'e structure (A1), \textit{fcc} and \textit{hcp} appear as the two less intense maxima (A2 and A3), exhibiting two different intensity levels. In contrast to graphene on Ru(0001)~\cite{Marchini:2007,Martoccia:2008,Sutter:2008,Praga:2008,Wang:2008}, in the present study we observe that bright features (local maxima) are surrounded by dark depressions (A4) present in both halves of the moir\'e unit cell [\ref{figure02}] corresponding to the "bridge" areas, which we believe are strongly bonded to the metallic substrate. The full graphene honeycomb lattice is only visible in areas directly on top or adjacent to the A1 maxima (moir\'e maxima) [~\ref{figure02} (b)] which we attribute to the weak coupling of these regions of the graphene sheet to the Rh lattice. The atomic structure of other regions is dominated by the strong covalent interaction with the metal substrate, especially within the black minima (A4) where C atoms are placed in "bridge" positions [see~\ref{figure02} (d)]. Recent theoretical investigations showed that for graphene on Ni(111)~\cite{Fuentes-Cabrera:2008} as well as for graphene on Ru(0001)~\cite{Moritz:2010} such "bridge" positions are  energetically favorable which might lead to a relatively strong C\,-\,Rh bonding in these regions. Especially the "bridge-top" position, which we believe gives rise to the observed black depressions in the STM images, was found to be one of the most energetically stable configurations~\cite{Fuentes-Cabrera:2008}.  Interestingly, the observed pronounced bonding of the "bridge-top" regions is almost completely suppressed in graphene on Ru(0001)~\cite{Marchini:2007,Martoccia:2008,Sutter:2008,Praga:2008,Wang:2008} where no local height variation within the half-cells of the moir\'e structure was observed. We also would like to point out that the atomic structure of graphene on Rh(111) also differs from those of \textit{h}-BN "nanomesh" on Rh(111)~\cite{Corso:2004,Dil:2008}.

\begin{figure}[t]
  \vspace{5mm}
    \begin{center}
\includegraphics[width=90mm]{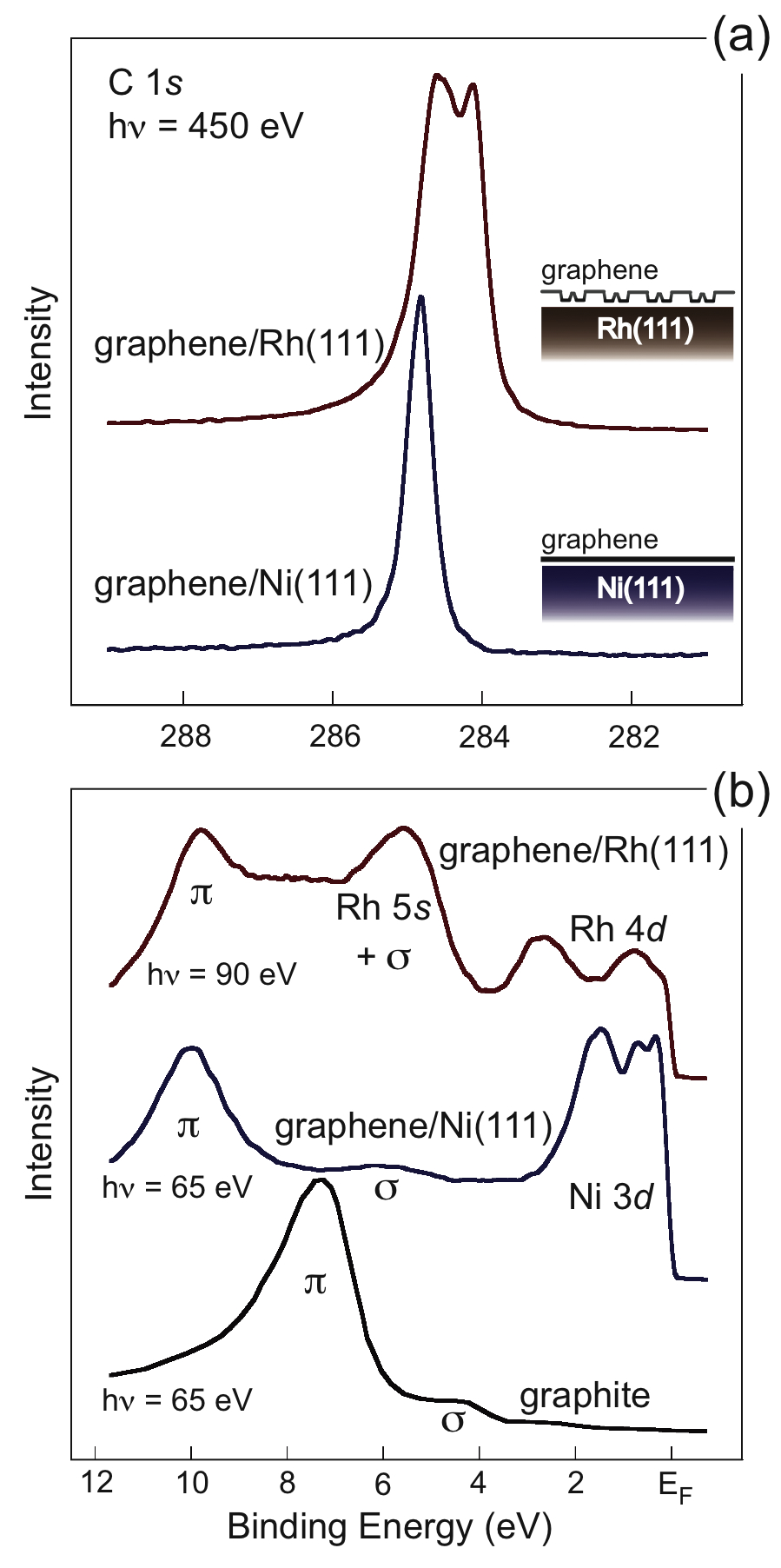}
    \end{center}
\caption{\label{figure03} (a)  C 1$s$ core level photoemission spectra of the graphene layer on Rh(111) and of the graphene layer on Ni(111) ($h\nu$\,=\,450\,eV). (b) Valence-band photoelectron spectra of graphene/Rh(111) ($h\nu$\,=\,90\,eV), graphene/Ni(111) ($h\nu$\,=\,65\,eV), and of graphite single crystal ($h\nu$\,=\,65\,eV) taken in normal emission geometry.}
\end{figure}

Core level and valence band PES provides strong support for the strong variations in bonding interaction between the carbon layer and the Rh(111) substrate. \ref{figure03} (a) shows the C 1$s$ photoemission spectra of graphene/Rh(111) and of graphene/Ni(111). In case of graphene/Rh(111) the line is split in to two distinct components with an energy separation of about 0.44\,eV. In an earlier study this double-peak structure was associated with a significant corrugation of the graphene layer on strongly bonding substrates~\cite{Preobrajenski:2008}. The component with lower binding energy is then due to the regions where graphene is weakly bonded to the substarte (A1), while the high-energy peak results from the strongly bonded regions (A2, A3 and especially A4). This assignment is confirmed by the absence of the lower energy feature in the strongly bonded, but relatively flat graphene layer on Ni(111) as well as by the interpretation of height differences in STM images of graphene/Rh(111) described above. \ref{figure03} (b) shows the valence band photoemission spectra of graphene/Rh(111),  graphene/Ni(111), and a pure graphite single crystal recorded in normal emission geometry. The difference in binding energy of the $\pi$ states in graphene/Rh(111) and pure graphite amounts to about 2.3\,eV, which is close to the value for graphene/Ni(111) and also for graphene/Ru(0001). In our view, this shift reflects the effect of hybridization of the graphene $\pi$ bands with the Rh 4$d$ bands and, to a lesser extent, with the Rh 5$s$ and 5$p$ states. These results indicate that the bonding strength of the graphene layer on top of the Rh(111) surface is comparable to that observed in graphene/Ru(0001). However, a considerable local redistribution of the orbital hybridization in graphene/Rh(111) compared to Ru(0001) or Ir(111) can be deduced from strong contrast variations observed by STM. Additionally we would like to compare the electronic structure of graphene and \textit{h}-BN and graphene grown on Rh(111). Previously, $\pi$ and $\sigma$ band splitting of \textit{h}-BN on Rh(111) was observed and assigned to the dielectric nature of \textit{h}-BN and the difference of local work functions between the "hole'' and the "wire'' regions~\cite{Dil:2008,Laskowski:2007}. For graphene/Rh(111) we observe no such splitting; this could be due to a smaller corrugation of the graphene layer or due to the metallic nature of graphene. Although both in $h$-BN/Rh(111) and in graphene/Rh(111) a strong vertical corrugation leads to the observation of two core-level lines, it turns out that the factors leading to the splitting are not sufficient to influence the valence band structure of $h$-BN and graphene layers in the same way.  These findings coincide with the recently reported results of the study of the valence-band structure of the graphene/Ru(0001) system, where no splitting of $\pi$ and $\sigma$ bands could be observed~\cite{Brugger:2009}.

In summary, we have grown epitaxial single layer graphene on the Rh(111) surface. High-resolution STM imaging shows that the graphene moir\'e structure has a distinctly different atomic configuration compared to those previously reported for graphene on Ir(111) or on Ru(0001) as well as for the \textit{h}-BN "nanomesh". The graphene supercell contains additional regions in both supercell subunits where the graphene layer is strongly bonded to the Rh(111) surface. These additional areas appear as depressions in STM images which is an indication for the strong orbital hybridization between the graphene layer and the metallic substrate. We attribute these regions to the "bridge" regions of the moir\'e structure. Valence-band photoemission confirms the strong hybridization between graphene an Rh(111) which leads to the pronounced corrugation of the graphene layer. The presented results emphasize the importance of considering substrate effects in epitaxial graphene/metal systems for a rational design of graphene-based nanoscale systems.

\textbf{Acknowledgement.} This work was supported by the Research Center "UltraQuantum" (Excellence Initiative) and by the Deutsche Forschungsgemeinschaft (DFG) via the Collaborative Research Center (SFB) 767. M. F. gratefully acknowledges financial support from the Baden-W\"urttemberg Stiftung.

\bibliography{fonin}


\end{document}